\documentclass[journal=apchd5,manuscript=article]{achemso}
\setkeys{acs}{articletitle = true}

\usepackage[version=3]{mhchem} 
\usepackage{xcolor} 
\usepackage{comment} 
\usepackage{siunitx} 

\author{R.~G.~Polozkov}
\author{A.~A.~Shubnic}
\author{I.~A.~Shelykh}%
\affiliation{Department of Physics and Engineering, ITMO University, St. Petersburg 197101, Russia}%
\alsoaffiliation{
Science Institute, University of Iceland, Dunhagi 3, IS-107, Reykjavik, Iceland
}
\author{I.~V.~Iorsh}%
\affiliation{Department of Physics and Engineering, ITMO University, St. Petersburg 197101, Russia}%
\email{i.iorsh@metalab.ifmo.ru}

\title[An \textsf{achemso} demo]
{High refractive index and extreme biaxial optical anisotropy of rhenium diselenide for applications in all-dielectric nanophotonics.}

\abbreviations{}

\begin{document}


%
%


\begin{abstract}
We establish simple quantitative criterium for the search of new dielectric materials with high values of the refractive index in the visible range. It is demonstrated, that for light frequencies below the band gap the latter is determined by the dimensionless parameter $\eta$ calculated as the ratio of the sum of the widths of conduction and valence bands and the band gap. Small values of this parameter, which can be achieved in materials with almost flat bands, lead to dramatic increase of the refractive index. 
 We illustrate this rule with a particular example of rhenium dichalcogenides, for which we perform ab-initio calculations of the band structure and optical susceptibility and predict the values of the refractive index $n>5$ in a wide frequency range around 1 eV together with compratively low losses. Our findings open new perspectives in search for the new high-index/low loss materials for all-dielectric nanophotonics.

\end{abstract}


\section{Introduction}

All-dielectric photonics~\cite{kuznetsov2016optically,kivshar2018all} is arguably the most rapidly evolving field of the modern nano-optics. The basic component in all-dielectric photonics, a dielectric nanoantenna, supports optical Mie resonances~\cite{mie1908beitrage}, with properties that can be flexibly controlled by the geometry of a nanoantenna. In particular, the variation of its shape allows to change the nature of the lowest energy optical resonance from electric dipole to magnetic dipole~\cite{kuznetsov2012magnetic}, the functionality inaccesible in plasmonic devices. With current stage of technology the fabrication of specifically designed arrays of resonant antennae with finely controlled shape and lattice geometry became a routine task, which paved the way towards unprecendented control over linear~\cite{decker2015high,arbabi2015dielectric,jahani2016all} and nonlinear~\cite{shcherbakov2015ultrafast,schlickriede2020nonlinear} light manipulation.

The key factor defining the functionality of resonant  all-dielectric nanostructures, besides their shape, is the value of the refractive index of the material forming resonant nanoantennas. Indeed, the quality factor of the fundamental Mie resonance scales as $n^2$ and the ratio of the nanoantenna size to the resonant wavelength as $1/n$. Thus, the increase of the refractive index would allow for both more sharp and more deep subwavelength resonances in all-dielectric nanostructures, provided that the extinction coefficient, related to the imaginary part of the refractive index, is kept small.

While virtually purely real refractive indices $n>10$ can be easily achieved in wide frequency range from THz to mid-IR, in the near-IR and visible ranges the values of refractive index are currently limited by a value of approximately 5. Around characteristic photon energy of 1 eV, standard high index reference materials are crystalline silicon ($n\approx 4$), gallium arsenide ($n\approx 3.7$) and germanium ($n\approx 4.5$). It is clear that the broadband large refractive index in this frequency range is provided by strong interband polarization. However, the questions of what are the other parameters which can contribute to high values of off-resonant refractive index, are there any fundamental upper limit for this quantity, and whether it can be substantially increased with respect to the values characterizing current high-index materials remain unexplored, up to our knowledge.

In this work we show that the sum of the widths of the conduction and valence bands to the bandgap, 
\begin{equation}
\eta=\frac{\Delta_c+\Delta_v}{E_G}    
\end{equation}
is the key parameter which defines the value of the refractive index for the frequencies slightly below the band gap. We derive a simplified estimation for the susceptibility, and demonstrate, that the refractive index can be substantially increased if $\eta$ becomes small, as it happens in materials with flattened valence and conduction bands. This situation takes place in ReSe$_2$, for which we perform the ab-initio analysis of the optical properties, and predict that for the photon energies around 1 eV the real part of the refractive index $n>5$, which is current record high value, and its imaginary part remains small. We also predict extreme biaxial optical anisotropy in this material.


\section{Theoretical methods}

We start from considering a simplified model of interband polarization in bulk material and show that the condition of the weak dependence of the band gap on the wave vector in the whole Brillouin zone (the extreme case being a material with flat bands) results in a high value of optical susceptibility close to the absorption edge.

We start from the simplest expression for the susceptibility tensor given by the Kubo formula

\begin{align}
    \chi_{\alpha,\beta}(\omega)=\frac{\hbar}{\omega\Omega N_k}\displaystyle\sum_{k,m,n}\frac{f(\epsilon_{nk})-f(\epsilon_{mk})}{\epsilon_{nk}-\epsilon_{mk}}\frac{\langle nk |j_{\alpha}|mk\rangle\langle mk | j_{\beta}|nk\rangle}{\hbar\omega-(\epsilon_{nk}-\epsilon_{mk})+i\gamma},
\end{align}
where  $\alpha,\beta=x,y,z$, $\Omega$ is the unit cell volume, $N_k$ is the number of points in the Brilloin zone, over which we make the summation, $f$ are the Fermi-Dirac distributions, $|nk\rangle$ are the Bloch eigenfunctions for the corresponding band and  $j_{\alpha}$ are the current operators given by:
\begin{align}
    \mathbf{\hat{j}}=-\frac{e}{\hbar} \nabla_{k}H_{k},
\end{align}
Where $H_k$ is the hamiltonian in the basis of Bloch eigenfunctions.

Let us consider the simplest case of a cubic unit cell, for which  $\Omega=D^3$, where $D$ is the lattice period. We limit ourselves to the case of single conduction and valence bands separated by the band gap $E_G$, assuming that  $E_G\gg kT$ and putting $f_c=0,f_v=1$. We then can rewrite the expression for the susceptibility as:
\begin{align}
    \chi_{\alpha,\beta}(\omega)=\frac{\alpha}{(k_0D)} \left(\frac{\hbar^2}{m_0 D^2}\right)^2\frac{1}{N_k}\sum_k \frac{D^2 P_{\alpha}^*(k)P_{\beta}(k)}{[\epsilon_c(k)-\epsilon_v(k)][\hbar\omega-(\epsilon_c(k)-\epsilon_v(k))+i\gamma]}
\end{align}
where $\alpha\approx 1/137$ is the fine structure constant, $k_0=\omega/c$, $m_0$ is the free electron mass,  $P_{\alpha}(k) = \int d^3 \mathbf{r} u_{c,k}^*(\mathbf{r}) i(\partial/\partial \mathbf{r}_{\alpha} u_{v,k}(\mathbf{r})$ is the matrix element of inter-band polarization. The presence of the prefactor $\alpha/(k_0 D)$ allows to understand, why one usually observes large permittivities at longer wavelengths. Indeed for the frequency $2\pi\hbar\omega = 1$ GHz and typical unit cell size $D\approx 5.7\times 10^{-10}$~m (lattice constant of germanium) the term $\alpha/(k_0 D)$ can be as large as $10^6$. on the other hand, in the range of optical frequencies $\hbar\omega\approx 1$ eV, $\alpha/(k_0 D)\approx 2.5$. The characteristic energy $\hbar^2/(m_0 D^2)\approx 0.5~\mathrm{eV}$ is comparable with the bandgap width.

At the first sight it seems, that the recipe for the large value of susceptibility is straightforward: if the material has $\epsilon_c(k)-\epsilon_v(k)\approx \mathrm{const}(k)=\tilde{\epsilon}$ then tuning the frequency slightly below $\tilde{\epsilon}$ we can achieve arbitrary large susceptibility with vanishing losses. Unfortunately, this would not work so directly, since conduction and valence band dispersions and matrix elements $P$  are directly related to each other. This can be easily seen if we use $\mathbf{k\cdot p}$ perturbation theory. As a very rough approximation we can assume that the dispersion in the whole Brilloin zone is parabolic and isotropic (which of course is not the case in real materials) and write 
\begin{align}
    \epsilon_c(k)-\epsilon_v(k)=E_G+\hbar^2k^2\left(\frac{1}{2m_c}-\frac{1}{2m_v}\right)=E_G+\frac{\hbar^2k^2}{2\mu}=E_G+\frac{\hbar^2 k^2}{2m_0}\left(\frac{4\hbar^2|P|^2}{m_0E_G}\right),
\end{align}
Where the last equality follows from two band $\mathbf{k\cdot p}$ perturbation theory. We can see that the constant energy difference between the two bands can be achieved only if  $|P|^2$ vanishes, but this means that optical transitions between valence and conduction bands are forbidden, which corresponds to zero net susceptibility. The dependence of $\chi$ on the bandwidths is therefore determined by the competition between the numerator and denominator in Eq. 3. 

If we set $\hbar\omega=E_G-\delta,\quad \delta\ll E_G$ and go from summation to the integration over the Brilloin zone, we can derive  simple analytical expression for susceptibility:
\begin{align}
    \chi(\omega)\approx \frac{\alpha}{(k_{E_G}D)}\frac{1}{\pi^2}\frac{\tan^{-1}(\pi\sqrt{\eta})-2^{-\frac{1}{2}}\tan^{-1}(\pi\sqrt{\eta/2})}{\sqrt{\eta}}, \label{chi_est}
\end{align}
where $k_{E_G}=E_G/(\hbar c)$,  $\eta=\hbar^2/(\mu D^2 E_G) = (\Delta_c+\Delta_v)/E_G$, where $\Delta_c$ and $\Delta_v$ are the bandwidths of the conduction and the valence bands, respectively. This simplified expression substantially underestimates the dielectric constant for real semiconductors (e.g. for GaAs it gives the estimate for $\varepsilon\approx 7$ instead of experimental value $\varepsilon\approx 13.5$. This is mainly due to the effects of non-parabolicity of the electron dispersion and neglect of the interband polarization corresponding to the higher bands. Nevertheless,  Eq.~\eqref{chi_est} allows to determine qualitatively the dependence of the susceptibility on $\eta$, as it contains the universal decaying function of this parameter.  Therefore, although matrix element of the interband transition decays with decrease of $\eta$, maximal values of the susceptibility should be still expected for materials with small $\eta$, i.e. for those having large gap and narrow conduction and valence bands. 

Fortunately, one can find representatives of this class of the materials, the example being bulk ReSe$_2$~\cite{Gunasekera}. This layered material belonging to the rhenium dichalcogenides family has  been gaining recently increasing attention, mainly due to its pronounced in-plane anisotropy and suppressed interllayer Van der Waals coupling~\cite{zhao2015interlayer}. To check the results of our qualitative analysis, in the next section we provide the data of ab-initio modelling of the linear optical response of bulk ReSe$_2$. 

\section{Results of the ab-initio modelling}

All ab-initio calculations were performed using QUANTUM
ESPRESSO package \cite{giannozzi09}. The analysis of the optical response was performed in three steps.

At the first step, we determined the equilibrium positions of the ions in the lattice by full self-consistent geometry optimization within DFT. The top and side views of the resulting structure are presented in Figure 1. The obtained lattice parameters and their comparison to experimental XRD data are given in the Supplementary. 

At the second step, we employed two different DFT approaches (LDA and GGA) to calculate the three-dimensional band structure.

At the third step we calculated the diagonal components of the unit cell polarizability  tensor $\alpha_{ii}$ via time dependent density functional theory (TDDFT) (see details in the Supplementary).

\begin{figure}[H]
\centering
\includegraphics[scale=0.45]{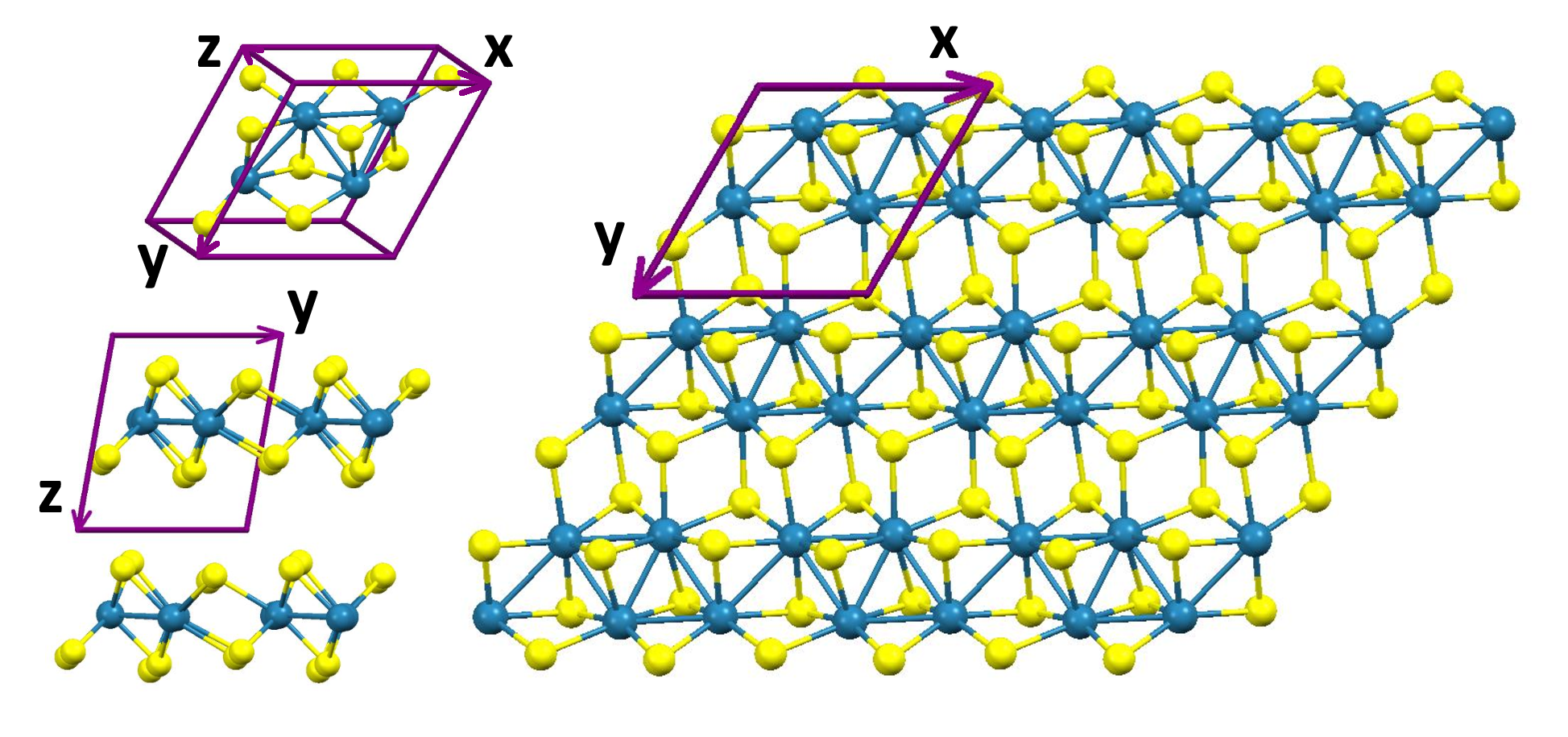}
\caption{The unit cell, side and top views, of the $ReX_{2}$ bulk structure. The x, y, z are crystallographic axes are marking the unit cell. Rhenium and sulfur (or selenium) atoms are indicated in blue and yellow, respectively. It can be seen that the set of the elementary cells forms a quasi-two-dimensional bulk material.}
\label{Unite sell:2x1x2}
\end{figure}


Most of the previous calculations of the band structure of the considered materials focused only on the dispersion along the special sets of directions, example being the 2D projection into the layers plane \cite{Gunas_24, Gunas_25}. In the other cases, highly symmetric paths were chosen in the 3D BZ connecting three-dimensional special points, without exploring the structure of the whole BZ \cite{Echeverry, Gunas_27} and thus missing the true band edges, which are crucially important for the determination of the optical response. The possible drawbacks of this approach were analyzed in detail in the recent work \cite{Gunasekera}, where the necessity of the calculation of the band energies everywhere within the BZ and determination of the constant energy surfaces was stressed.


In the present paper we used the k-path, developed in Ref.\cite{Gunasekera}, and calculated the band structure using the following path: $\Gamma$ - CBM - CBM' - Z - $\Gamma$ - VBM - VBM' - Z - $\Gamma$, where CBM - Conduction Band Minimum (LUMO), VBM - Valence Band Maximum (HOMO) and CBM', VBM' are the intersection points of the boundary surface of the Brillouin zone and the straight lines connecting $\Gamma$ and CBM or VBM points respectively. The paths from CBM' or VBM' points to the Z point run along the surface of the BZ containing Z point.  

To find the k-points corresponding to the VBM and CBM we used the different grids in BZ - 6x6x6 and 7x7x7. The calculated values of the direct and indirect band gap are presented in Table \ref{table2} and compared with previously reported data \cite{Gunasekera, Echeverry}.  

\begin{table}
\centering
\caption{Band gap values calculated within LDA and GGA. The grid 6x6x6  was used to define the VBM point for ReSe2 and CBM point for ReS2, and the grid 7x7x7 was used to define the CBM point for ReSe2 and VBM point for ReS2. This choice was dictated by the reasons of grid parity and symmetry of the Brillouin zones. All energies are given in eV.}
\begin{tabular}{|c|c|c|c|}
\hline
$ $ & LDA/GGA & LDA \cite{Gunasekera}/GGA \cite{Gunasekera} & GGA+GW \cite{Echeverry} \\ \hline
indirect $ReSe_2$ & 1.0043/1.0463 & 0.87/0.99 & -   \\  \hline 
direct (Z) $ReSe_2$ & 1.0714/1.0909 & 0.97/1.00 & 1.38  \\   \hline
direct ($\Gamma$) $ReSe_2$ & 1.3262/1.3421 & - & -  \\   \hline
indirect $ReS_2$ & 1.1424/1.1790 & - & -   \\  \hline 
direct (Z) $ReS_2$ & 1.1908/1.2147 & - & 1.60  \\   \hline
direct ($\Gamma$) $ReS_2$ & 1.6809/1.7441 & - & 1.88  \\   \hline
\end{tabular}
\label{table2}
\end{table}

The obtained band structures of $ReSe_2$ and $ReS_2$ are presented in Figure 2. It can be seen, that the energy of the lowest conduction and the highest valence bands  depend only weakly on the wavevector, and the bands are close to the flat ones.

\newpage
\begin{figure}[H]
\centering
\includegraphics[scale=0.3]{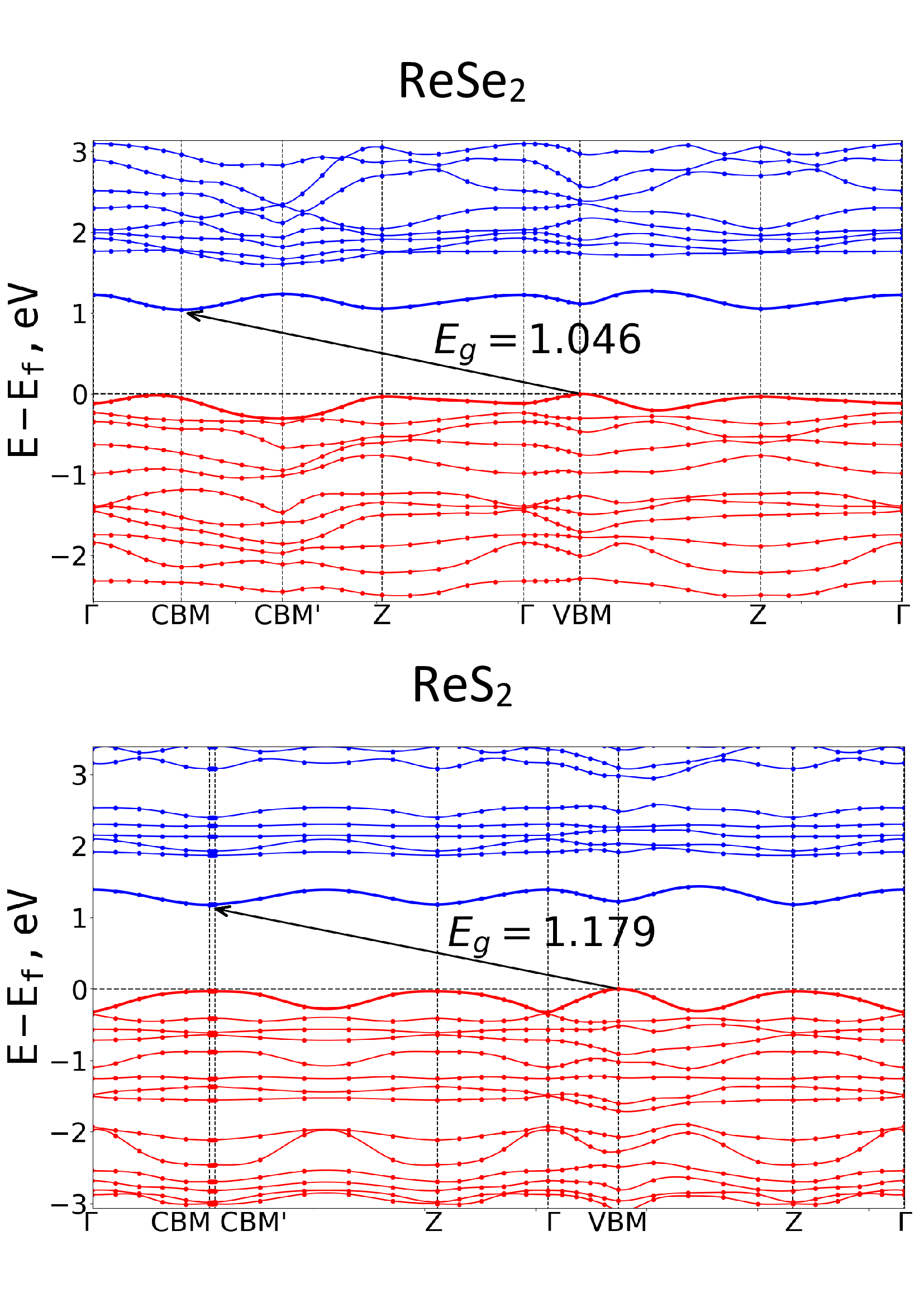}
\caption{The band structure of the bulk $ReSe_{2}$ and $ReS_{2}$ calculated by GGA level of DFT for the path going through $\Gamma$ and $Z$ points and the band extrema (CBM and VBM). The Fermi energy is set to zero. The CBM' and VBM' points are the intersection points of the boundary surface of the Brillouin zone and the straight lines connecting the $\Gamma$ point and CBM or VBM points respectively.
The bands are flattened in a wide range of the wave vectors. CBM and VBM points correspond to the different k-values so this is indirect semiconductor.}
\label{Bands-GGA}
\end{figure}


\subsection{Optical response}

\begin{figure}[H]
\centering
\includegraphics[scale=0.6]{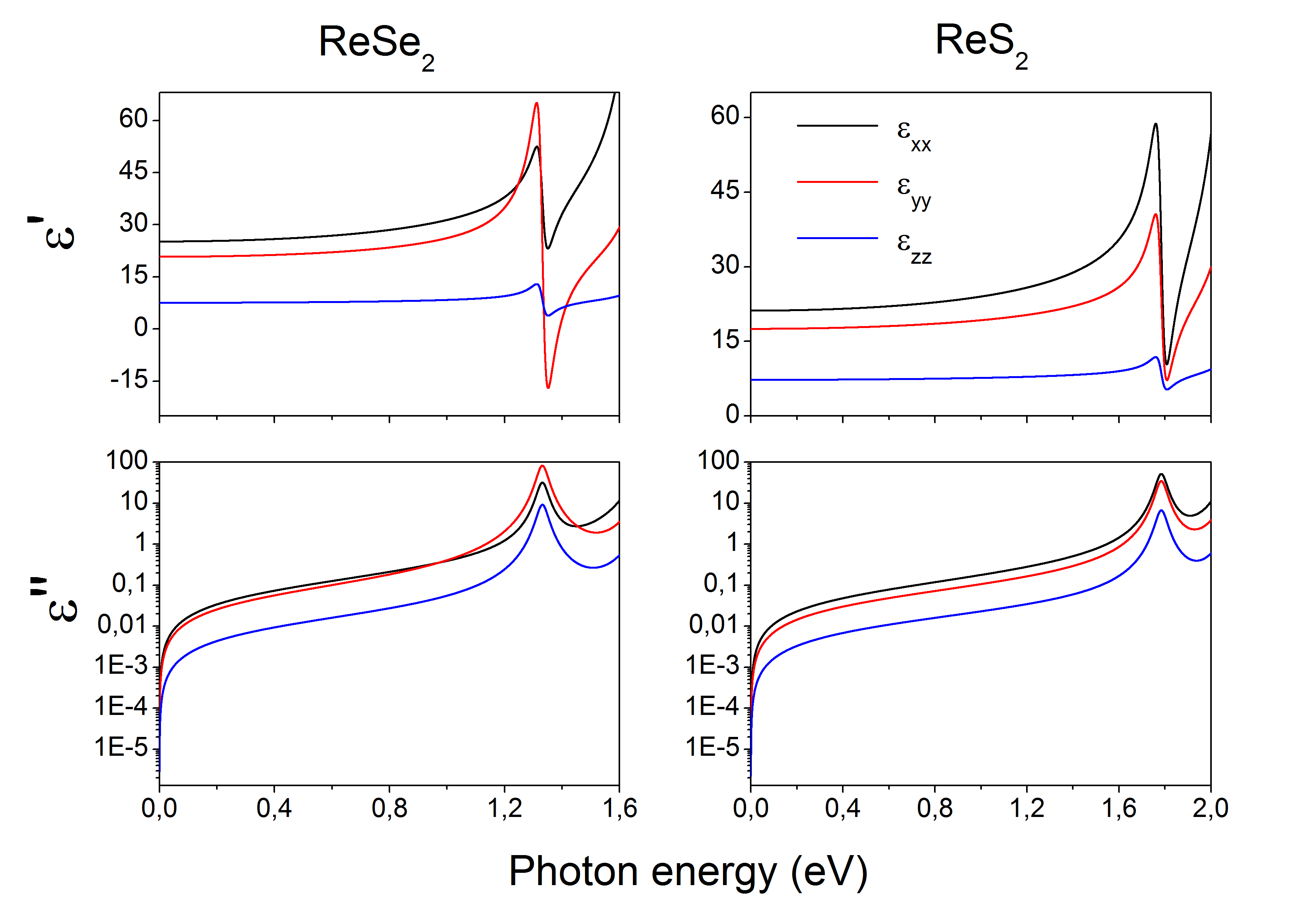}
\caption{Real and imaginary parts of $\varepsilon$ of the bulk $ReSe_{2}$ (left) and $ReS_{2}$ (right) for different light polarizations calculated within TDDFT (see details of calculations in Supplementary). It can be seen that the peaks of the imaginary part of the permittivity, and hence the absorption edges, are 1.32 eV for ReSe$_2$ and 1.7 eV for ReS$_2$.}
\label{epslion_xx_yy_zz_ReSe2}
\end{figure}

In order to get the the optical response we employ the TDDFT approach, calculating the diagonal terms of the unit cell polarizability $\alpha_{ii}$. The average dielectric susceptibility $\chi_{ii}(\omega)$ was found as the ratio of the polarization and the unit cell volume $V$, $\chi_{ii}=\alpha_{ii}/V$. The dielectric permittivity was then found as $\varepsilon_{ii}=1+4\pi\chi_{ii}$. The spectra of the dielectric permittivities are shown in Fig.~\ref{epslion_xx_yy_zz_ReSe2}.

\begin{figure}[H]
\centering
\includegraphics[scale=0.6]{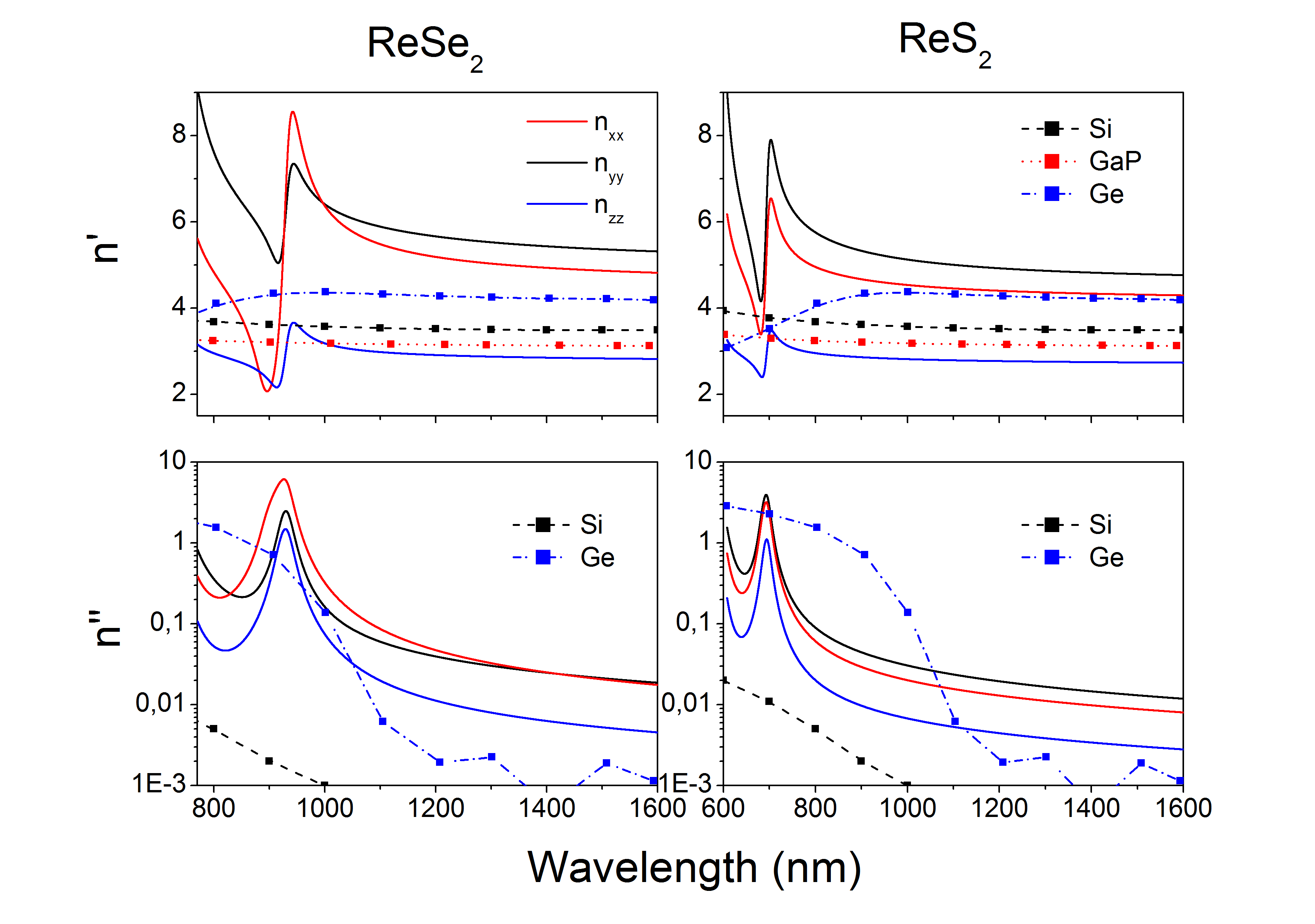}
\caption{
The calculated dependence of the the real and imaginary parts of refractive indexes of $ReSe_{2}$ (left) and $ReS_{2}$ (right) on the wavelength for different cases of light polarizations. The known experimental data for the modern state of the art high-index materials such as c-Si~\cite{Si}, GaP~\cite{GaP} and Ge~\cite{jellison1992optical} are shown for comparison.
}
\label{Refr_rese2}
\end{figure}

It can be seen that for both ReSe$_2$ and ReS$_2$ the absorption sharply increases only when the photon frequency approaches the band gap in $\Gamma$ point, 1.32 and 1.7 eV, respectively. Furthermore, we observe strong biaxial anisotropy for both crystals, stemming both  from the layered structure of the material and reduced in-plane symmetry of the rhenium dichalcogenides~\cite{tian2016low}. Noteworthy, large refractive index contrast and strong uniaxial anisotropy has been recently experimentally measured in bulk transition metal dichalcogenides~\cite{verre2019transition,busschaert2020tmdc,ermolaev2020giant} and hexagonal boron nitride~\cite{Segura2018}. The pronounced biaxial anisotropy may be used for the the observation of the Dyakonov surface waves~\cite{d1988new} in the rhenium dichalcogenide waveguides.
One can see that the real parts of the in-plane components of the permittivity of ReSe$_2$ exceed 25 in broad frequency range, which should correspond to high refractive index. Since for biaxial crystals the refractive index depends on the light propagation direction, we limit ourselves to the case when light travels along one of the three principal axes. In this case, the refractive index for each of the two polarizations is just a square root of the corresponding permittivity. The spectra of the refractive index components is shown in Fig.~\ref{Refr_rese2}. One sees that in broad 
wavelengths range the refractive index exceeds 5 which is substantially larger that the index of the modern state of the art high-index materials such as c-Si~\cite{Si}, GaP~\cite{GaP} and Ge~\cite{jellison1992optical}. The refractive index of ReS$_2$ is smaller than that for ReSe$_2$, but is still quite large. It should be noted that the imaginary part of the refractive index in this 
wavelengths range is negligibly small. 

In our analysis, we have not accounted for the excitonic contribution to the dielectric permittivity, which can be substantial and would lead to the additional absorption losses at the frequencies slightly below the band gap. Nevertheless, since the range of large refractive index covers more than 0.5 eV below the band gap, it should be possible to detune from the exciton absorption lines to probe the lossless large refractive index.

\textit{Conclusions}
We established simple quantitative criterium for the search of high refrective index dielectric materials, expressed in terms of the single dimensionless parameter calculated as the ratio of the sum of the widths of conduction and valence bands and the bandgap. With the use of it, we found that ReSe$_2$ is low loss material, possessing record high refractive index $n>5$ in a wide frequency range around 1 eV. Our conclusion is supported by the results of the DFT simulation of the optical response.  Since currently there exist wast material databases containing the data on their bandstructures, the proposed criterion can be used for the automated search of the perspective candidates for the novel high-index materials for all-dielectric photonics.

\section{Acknowledgement}
This work was supported by the Ministry of Science and Higher Education of Russian Federation, goszadanie no. 2019-1246.
The authors acknowledge the support from the mega-grant No. 14.Y26.31.0015 of the Ministry
of Education and Science of the Russian Federation. I.A.S. acknowledges the support from the Icelandic research fund, grant No. 163082-051.

\newpage
\begin{suppinfo}
This material is available free of charge via the Internet at http://pubs.acs.org.


%
\end{suppinfo}

\bibliography{bibl}


\end{document}


\begin{suppinfo}

\section{Computational details of the ab-initio modelling}\label{A1}

We performed three main computational steps. The first one was a full optimization of the spatial structure within DFT for the relaxation of a stressed system and determination of the equilibrium configuration. We used X-ray diffraction data \cite{XRD} as the starting coordinates and Broyden-Fletcher-Goldfarb-Shanno (BFGS) 
quasi-newton algorithm for geometry optimization. We used fully relativistic pseudopotentials with the projector augmented wave (PAW) method \cite{PAW} from PSLibrary \cite{PAW_base}. Calculations were performed with convergence threshold on the total energy for ionic minimization $10^{-6}$ (a.u) and convergence threshold on forces for ionic minimization $10^{-5}$ (a.u). The Monkhrost-Pack grid 6x6x6 was used for geometry optimization.
Kinetic energy cut-off corresponds to 60 Rydberg for wave functions and 300 Rydberg for charge density and potential with convergence threshold for self consistency $10^{-10}$. The results of the geometry optimization and their comaparison to the experimental data is summarized in Tabel 1.
\begin{table}
\centering
\caption{Calculated and experimental lattice parameters and angles for bulk $ReX_2$ systems}
\begin{tabular}{|c|c|c|c|c|c|c|}
\hline
& x(\AA) & y(\AA) & z(\AA) & $\alpha$ (\deg) & $\beta$ (\deg) & $\gamma$ (\deg) \\ 
\hline
$ReS_2$ &  &  &  &  &  &  \\ 
\hline
Exp bulk \cite{ExpBulk1} & 6.510 & 6.461 & 6.417 & 88.38 & 106.47 & 121.10 \\ \hline
DFT bulk & 6.310 & 6.414 & 6.301 & 91.59 & 105.27 & 118.78 \\ \hline
$ReSe_2$ &  &  &  &  &  &  \\
\hline
Exp bulk \cite{ExpBulk2} & 6.602 & 6.716 & 6.728 & 91.82 & 104.90 & 118.94 \\ \hline
DFT bulk & 6.597 & 6.710 & 6.721 & 91.84 & 104.90 & 118.91 \\ \hline
\end{tabular}
\label{table1}
\end{table}


The second step was a self-consistent calculation of electronic structure within DFT by pw.x code as a part of QE package. This code allows to calculate ground-state energy and one-electron Kohn-Sham orbitals. For ground state the number of computed bands was equal to number of electron in the unit cell. 
A band structure calculation was provided by bands.x code of QE package. The number of occupied and unoccupied orbitals was equal  188 and 72 for $ReSe_2$ and 108 and 42 for $ReSe_2$ respectively. As it was shown in the recent work \cite{Gunasekera} only generalized gradient approximation (GGA) instead of local density approximation (LDA) is convenient for band structure calculations of these systems. To check it we computed self-consistent calculation by two levels of DFT - LDA and GGA, and compared obtained results between themselves and with work \cite{Gunasekera}. The Perdew–Zunger \cite{PW_LDA} for the local density approximation (LDA) and Perdew–Burke–Ernzerhof \cite{pbe96} for generalized gradient approximation (GGA) exchange–correlation functionals are used.
The bands was calculated for different choices of grids to cover the whole 3D Brilloin zone. This process is not usually
required in more high-symmetry materials and
can be computationally expensive. But as it was shown in work \cite{Gunasekera} discarding the third component of the wave vector for the bulk rhenium dichalcogenides leads to impossibility to determine whether the band extrema occur at the same crystal momentum values.


The third step was a calculation of the optical response. To describe the absorption spectra we used the linear response time dependent density functional theory (TDDFT) based on the solution of quantum Liouville equation with PBE exchange-correlation kernel as well. It allows us to calculate absorption spectra of molecules using the Lanczos recursion algorithm without computing empty states \cite{TDDFT}. For this purpose we used turbo-lanczos.x program as a part of QE package. Physical quantities under study were the real and imaginary parts of the all component of the unit cell polarizability tensor $\alpha_{xx}$, $\alpha_{yy}$ and $\alpha_{zz}$ at the $\Gamma$ point. The number of Lanczos iterations was 20000 and the broadening/damping parameter was equal $25$meV. All data was interpolated by cubic B-splines.




\end{suppinfo}

\bibliographystyle{apchd5}
\bibliography{bibl}